\documentclass[%
aip,
 amsmath,amssymb,
preprint,%
]{revtex4-1}

\usepackage{graphicx}
\usepackage{dcolumn}
\usepackage{bm}
\usepackage{siunitx}

\usepackage[utf8]{inputenc}
\usepackage[T1]{fontenc}
\usepackage{mathptmx}
\begin{document}

\title{Precise, Subnanosecond, and High-Voltage Switching of Complex Loads Enabled by Gallium Nitride Electronics}

\author{John W. Simonaitis}
 \email{johnsimo@mit.edu}
 \affiliation{Research Laboratory of Electronics, Massachusetts Institute of Technology
}%
\author{Benjamin Slayton}%
\affiliation{%
Wentworth Institute of Technology
}%

\author{Yugu Yang-Keathley}
 \affiliation{%
 Wentworth Institute of Technology
}%

\author{Phillip D. Keathley}
 \affiliation{Research Laboratory of Electronics, Massachusetts Institute of Technology
}%

\author{Karl K. Berggren}
 \affiliation{Research Laboratory of Electronics, Massachusetts Institute of Technology
}%

\date{\today}

\begin{abstract}
In this work, we report the use of commercial Gallium Nitride (GaN) power electronics to precisely switch complex distributed loads, such as electron lenses and deflectors, without impedance matching. Depending on the chosen GaN field effect transistor (GaNFET) and driver, these GaN pulsers are capable of generating pulses ranging from 100 - 650 V and 5 - 60 A in 0.25 - 8 ns using simple designs with easy control, few-nanosecond propagation delays, and MHz repetition rates. We experimentally demonstrate a simple 250 ps, 100 V pulser measured by a directly coupled 2 GHz oscilloscope. By introducing resistive dampening, we can eliminate ringing to allow for precise 100 V transitions that complete a -10~V to -90~V transition in 1.5~ns, limited primarily by the inductance of the oscilloscope measurement path. The performance of the pulser attached to various load structures is simulated, demonstrating the possibility of even faster switching of internal fields in these loads. These circuits also have \SI{0.25}{\square\centi\meter} active regions and under 1~W power dissipation, enabling their integration into a wide variety of environments and apparatus. The proximity of the GaNFETs to the load due to this integration minimizes parasitic quantities that slow switching as well as remove the need to match from \SI{50}{\ohm} lines by allowing for a lumped element approximation small loads. We expect these GaN pulsers to have broad application in fields such as optics, nuclear sciences, charged particle optics, and atomic physics that require nanosecond, high-voltage transitions.

\end{abstract}

\maketitle

\section{\label{sec:Introduction}Introduction}
Nanosecond, high-voltage pulses have broad application in in the physical sciences, ranging from their use in optics for driving Pockels cells and piezoelectric actuators \cite{Bishop2006SubnanosecondTransistors}, to their use in deflecting and gating electrons or ions in nuclear science, spectroscopy \cite{Wolff1953ADispersion}, charged particle optics \cite{Zhang2020BeamDeflection, Zajfman1997ElectrostaticBeams}, and quantum measurement schemes \cite{Kruit2016DesignsMicroscope}. In the past, these fast pulses have been generated by a wide variety of technologies such as silicon power FETs \cite{Jiang2007FastMOSFETs, Chaney2004SimpleCircuit, Baker1992AMOSFETs}, avalanche transistor circuits \cite{Jinyuan1998HighTransistor, Henebry1961AvalancheCircuits, Benzel19851000-VDevices}, step recovery diodes \cite{Zou2017ADiodes}, non-linear transmission lines \cite{Afshari2005NonlinearSilicon}, spark gaps \cite{CarlE.BaumUltra-WidebandElectromagnetics}, and laser-triggered semiconductor gaps \cite{Mourou1979High-powerPrecision, Kohler2013AGeneratorb}. Various reviews and studies exist comparing some of these techniques  \cite{DeAngelis2011ExperimentalGenerators, Martin1992NanosecondTechniques, Mankowski2000ATechnology}, and most recently the use of nanoplasmas  \cite{SamizadehNikoo2020Nanoplasma-enabledElectronics} has given record switching performance. While some techniques deliver pulses greater than 100 kV in < 100 ps, these approaches have varied trade-offs and shortcomings. Among the most important are the cost and complexity of pulser designs, their large size and high power dissipation that force us place pulsers far away from the loads they drive, the lack of simple, single-shot driving schemes, and slow repetition rates. All of these techniques also have problems with ringing due to resonances in the pulse generating circuits and loads, and impedance mismatch problems developed over the length scales of the transmission line and loads.  For structures that requires precise switching, removing ringing requires either slowing of the pulse edge with low pass filtering to avoid load resonances, the use of resonant filters to selectively remove those worst resonances, or predistortion calibration techniques that require high bandwidth arbitrary waveform generators to create messy pulses that result in clean output switching \cite{Klopfer2020RFMirrors}.

In this work, we demonstrate the potential of simple and low-cost GaN power electronics for fast and precise high-voltage switching. While such electronics have been used extensively for high-efficiency power converters \cite{Wu2008AMHz}, amplifiers \cite{Brown2011W-bandMMICs}, and pulsed lasers \cite{Glaser2018HighFet}, we believe that these circuits have the potential for wide application in  physics and engineering for high power nanosecond switching. These circuits are fast and simple to control, with single-shot pulsing triggered by a 5 V logic inputs up to a repetition rate of 60 MHz and propagation delays of a few nanoseconds. The high power efficiency and low thermal dissipation of these circuits also offer advantages in operation, leading to small form factors that are compatible with vacuum environments common in physics applications. They are also inexpensive, costing two orders-of-magnitude less than equivalent commercial pulsers. Through careful selection of parts and layout optimization, the reported pulser achieves both undamped 250 ps and damped 1.5 ns, 100 V transitions (as measured by the time to go from -10~V to -90~V, or 10\% to 90\%) into a 2 GHz oscilloscope. These measurements are limited primarily by the path inductance into the  oscilloscope. We then simulate the response of various predominantly inductive, capacitive, and mixed loads to the GaNFET pulser, demonstrating precise nanosecond transitions even faster than those we measured experimentally, with ringing controlled to within 1\% the transition amplitude by resistive damping. We discuss alternative topologies allowing for symmetric pulse edges and higher voltage operation. Finally we discuss how our GaNFET-based switching performance in real loads compares to that of low-pass filtered pulses on \SI{50}{\ohm} lines. %

\section{Experimental Validation}
Wide bandgap semiconductors have long been considered excellent candidates for power electronics due to their ability to withstand voltage and currents far beyond that of silicon, low channel resistances, and high temperature compatibility \cite{Amano_2018_GaNRoadmap}. In recent years, GaN on silicon technology has emerged as a low cost and robust option with commercial products that have small gate and output capacitances, optimized packages with minimal parasitic inductances, and a wide variety of current and voltage ratings \cite{Lidow2019GaNConversion}. Along with these advances, commercial drivers with subnanosecond rise times, 7 A of peak drive current, 60 MHz repetition rates, and 2.5 ns propagation delays have emerged \cite{instruments2018lmg1020}. Together, these offer the possibility of subnanosecond switching of hundreds of volts and tens of amperes.

\begin{figure} [h]
    \centering
    \includegraphics[width=0.99\textwidth]{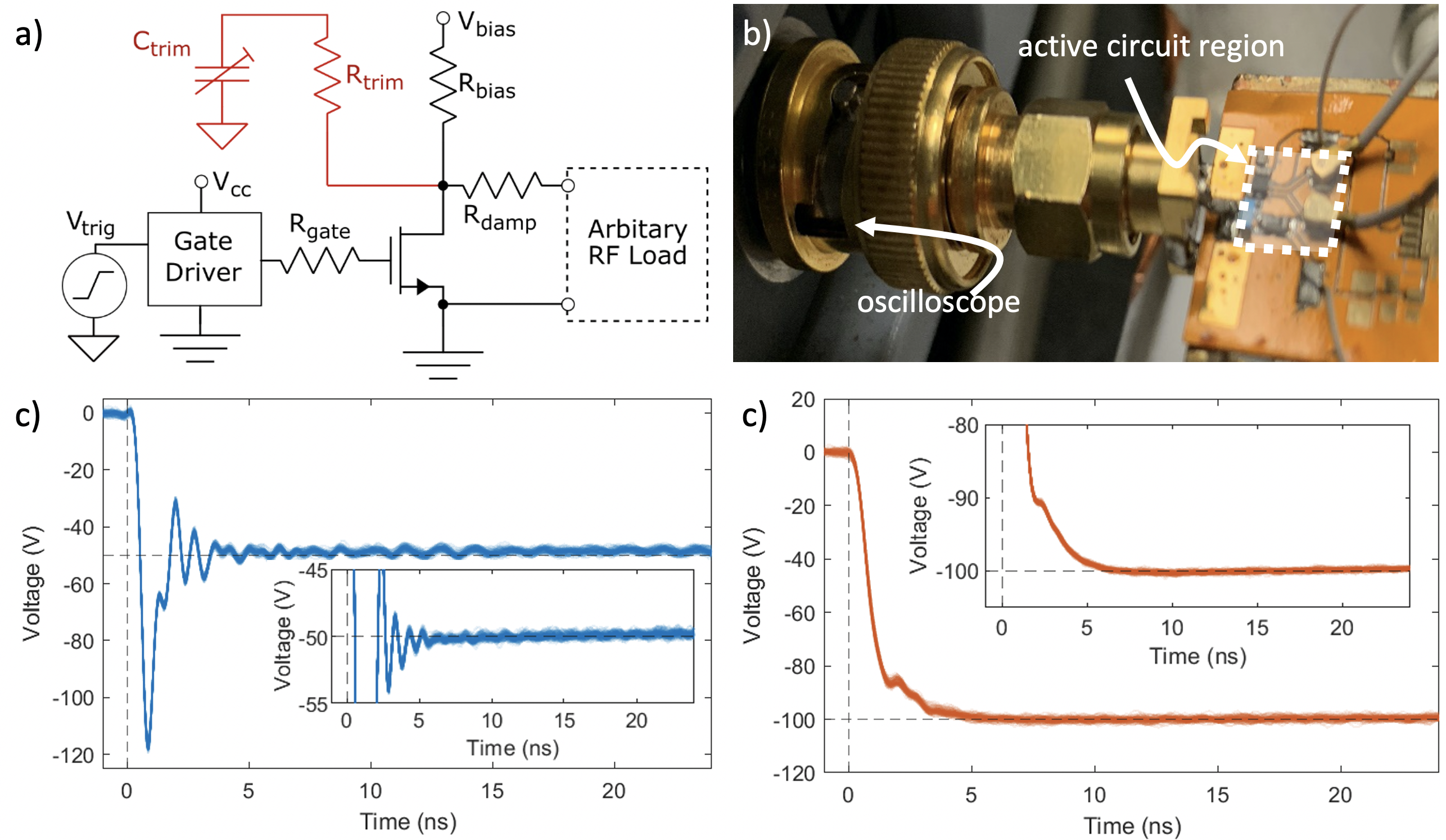}
    \caption{Test-bed for GaN pulsing. (a) The circuit schematic, where the gate driver is the LMG1020, and GaNFET the EPC2012C. ${}{R_{gate}}$ damps the gate turn on loop, ${}{R_{damp}}$ damps the load loop, and ${}{R_{bias}}$ sets the power usage and load recovery time. The optional trimming circuit in red allows for fine tuning of the load damping that the course-valued resistors used could not achieve, though this was not used in the the results shown. (b) The physical layout of the circuit mounted on a Kapton on copper substrate. The highlighted white region shows the active portion of the circuit which is less than \SI{ 0.25}{\square\centi\meter}. The circuit is directly coupled into a LeCroy 6200A oscilloscope with $\sim$20~pF of input capacitance through a 2.2~pF capacitor. (c) The undamped response, showing significant long-term ringing. The inset shows a zoomed in version of this ringing. The full transition is approximately 540~ps, while the 10\% to 90\% transition is in 230~ps. (d) The same pulse edge, but damped through a \SI{200}{\ohm} resistor. It achieves a 10\% to 90\% transition in 1.5~ns, 0\% to 90\% accuracy in 1.9~ns, and 0\% to 99\% accuracy in 5.1~ns. Note that all traces are multiplied by a factor of $\sim$10 to account for the voltage division caused by the 2.2~pF coupling capacitor, in order to match the final voltage to the biasing applied.}
    \label{fig:1}
\end{figure}

In order to validate these switching speeds, we built a testing circuit from the LMG1020 driver from Texas Instruments along with the 200~V EPC2012C enhancement mode GaNFET from the Efficient Power Corporation. The basic circuit design is shown in Figure 1a. The load is held to potential ${}{V_{bias}}$ until the gate driver is triggered by ${}{V_{trig}}$. When the driver is triggered, the gate driver outputs a high current signal through ${}{R_{gate}}$ to turn on the GaNFET by charging its gate. Charge is then shunted out from both the GaNFET’s parasitic output capacitance and the arbitrary load through ${}{R_{damp}}$ into ground, bringing the load bias to zero . When the GaNFET is shut back off, ${}{V_{bias}}$ pulls the voltage back up through ${}{R_{bias}}$, though this transition is significantly slower due to the larger value of ${}{R_{bias}}$ set to limit power dissipation in this element. The resistors ${}{R_{gate}}$ and ${}{R_{damp}}$ are also used to set the turn on time and damp and ringing for the FET and the load, respectively. A tunable trimming circuit, as shown in red in Figure~\ref{fig:1}a, was incorporated to ensure that the transition was critically damped, since the RF damping resistors used had values too coarse to precisely damp the circuit.

Figure~\ref{fig:1}b shows the realization of this circuit on a \SI{100}{\micro\meter} thick Kapton dielectric and copper substrate, directly coupled through an SMA and SMA-to-BNC adapter to the oscilloscope. The white shaded region shows the active area of the GaNFET, driver, bypass capacitors and damping, taking up less than \SI{0.25}{\square\centi\meter}, while the power supply makes up the rest of the circuit and can be placed off-board. Detailed optimization of the components and layout to minimize parasitic quantities and maximize switching speed is discussed in the supplement.

The undamped and damped response of the circuit coupled through a 2.2~pF capacitor to a 2 GHz LeCroy 6200A oscilloscope is shown in Figure 1c and d respectively. Initially we took measurements with a passive, 500 MHz, \SI{10}{\mega\ohm} probe (LeCroy PP007-WR), but we found that the distributed nature of the probe complicated our attempts to reduce the circuit to a simple lumped element model. This is discussed further in the supplement. The undamped response at ${}{R_{damp}}$ = 0 $\Omega$ shows a full, 0\% to 100\% transition in 540 ps, but suffers from massive overshoot and significant ringing. To ensure we did not damage the oscilloscope with this overshoot, we kept this measurement to 50~V, though when measured by the probe we easily pushed this to 100~V as seen in Figure S4b in the supplement. The damped response obtained at ${}{R_{damp}}$ = 200 $\Omega$ shows a transition from -10~V to -90~V (10\% to 90\% of the amplitude) in less than 1.5 ns with no residual ringing, as shown in Figure 1c. More significantly, the damped circuit is capable of reaching 10\% of the final voltage in under 1.9~ns, and 1\% accuracy in $\sim$5.1~ns as shown in the inset of Figure \ref{fig:1}d. We note that these circuits were successfully driven up to 200 volts without failure. However to ensure we did not damage the probe and oscilloscope, the maximum voltage at which we measured the transition was 100~V, as shown.

From these measurements, we can then estimated the parasitic quantities of our circuit. This assumes we can treat the whole system as lumped elements, which as discussed later is valid due to the small length scale of the GaNFET and load. Modeling the circuit as an series resistor, inductor, capacitor (RLC) circuit, as shown in Figure S5 in the supplement, we estimate the series capacitance to be 2~pF. This is calculated by putting the $\sim$20~pF oscilloscope capacitance in parallel with the 2.2~pF capacitor, and is validated by the 1:10 attenuated voltage transition we measure on the oscilloscope. 

We then measure the ringing frequency of the undamped load. This is approximately 1.1~GHz. Assuming that this measurement is relatively undamped, we can then use Eqn. 1 below to estimate the loop inductance. Here $\mathrm{L_{osc}}$ is the oscilloscope inductance, $\mathrm{f_{ring}}$ the ringing frequency, and $\mathrm{C_{osc}}$ the oscilloscope capacitance see  through the 2.2~pF coupling capacitor. This leads us to estimate the inductance to be $\sim$10~nH, which is roughly consistent with the \SI{200}{\ohm} damping we required to prevent ringing. This is also validated by the fact that commonly used RG-58/U coaxial cable has an inductance of 3.66~nH/cm, and the measured connector to oscilloscope length is $\sim$3~cm, meaning we would expect the cabling to give us $\sim$11~nH of inductance.

\begin{equation}
L_{osc} = \frac{1}{(2 \pi f_{ring})^2 C_{osc}}.
\end{equation} \label{eqn:inductorEst}


The response times of both of these transitions are limited by the inductance of the oscilloscope path, which results in a $\sim$1.1~GHz resonance. If we wanted an even faster measurement, we could try to reduce the loop inductance of the measurement by putting the GaNFET even closer to the oscilloscope. We could also use active probing, though these probes generally have voltage ratings too low for such high power switches and still perturb the circuit greater than many loads we are interested in. Furthermore, we are most interested in the fields generated in the loads of interest, not the voltage at different conductors, which probing can never directly tell us.  Thus, we used RF simulations to estimate the more fundamental field switching capabilities of these circuits important to many real-world applications. 

\section{\label{sec:theory}Simulation of Response to Realistic Loads}
In order to validate the performance of the GaNFET pulser for switching realistic loads, we simulated the GaNFET pulser’s switching performance in various systems, including a primarily capacitive load, a primarily inductive load, and a load with characteristics of both, as shown in Figure 2a, Figure \ref{fig:3}d, and Figure \ref{fig:3}a respectively.  

\begin{figure*} 
    \centering
    \includegraphics[width=1\textwidth]{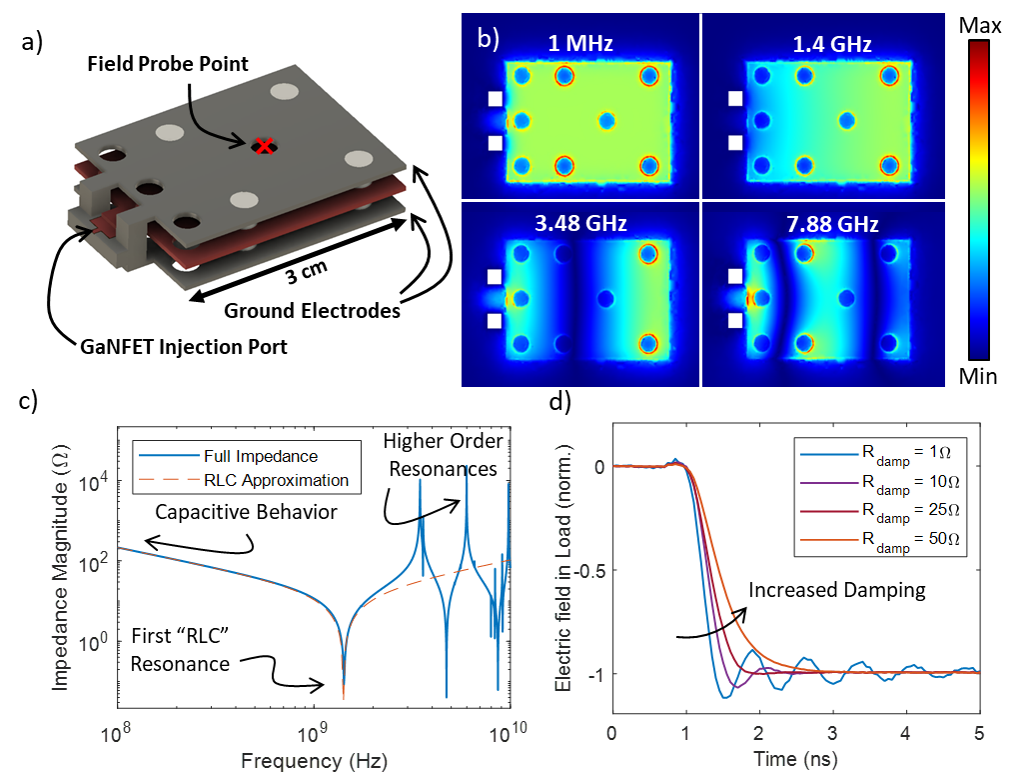}
    \caption{Simulated models of a parallel plate load. (a) The geometry of the load where the length of the plate is 3 cm, with the red x denoting the point at which we are probing the electric field. (b) The field structure of the load at various frequencies. At 10 MHz, the load all is driven in unison. Until $\sim$1 GHz, the field distribution is flat and behaves as a single lumped element. At 1.4 GHz, this begins to break down and we see the lowest mode develop. At 3.48 GHz and 7.88 GHz we see higher order modes develop, and the lumped element approximation completely breaks down. (c) The impedance that the driving GaNFET sees. At frequencies below 700 MHz it behaves almost as an ideal capacitor. Up to 2 GHz, a simple RLC circuit model works extremely well, as shown with the overlaid fit. In our simulations, we used rational polynomial based circuit models to fit to all of the resonances up to 10 GHz. We discuss this further in the supplement. (d) The damping of the response by setting $\mathrm{R_{damp}}$, which slows the response as it removes ringing. Note that higher-frequency ringing does still exist in the load fields due to the higher-order resonances, but they are less than 1\% the transition magnitude}
    
    \label{fig:2}
\end{figure*}

The basic approach and results of these simulations are outlined in Figure 2a-d. First, we defined a geometric structure in COMSOL Multiphysics to be excited. In the case of the primarily capacitive load, we defined three 304 stainless steel plates from Kimball Physics stacked with alumina spacers, with the central electrode (red) receiving the GaNFET excitation as shown in Figure \ref{fig:2}a. This kind of structure is widely used in electron optics for the fast gating (if the beam enters vertically) and deflection (if the beam enters horizontally) of charged particles. We defined the source and drain of the GaNFET to be our input ports for the simulations. 

Using the RF Module in COMSOL, we ran a full wave simulation of the structure in the frequency domain, extracting both the field distributions in the load and the port impedance seen by the GaNFET, as shown in Figure \ref{fig:2}b and \ref{fig:2}c respectively. The rational fit and the circuit model of the port impedance was generated in MATLAB and imported into LTSpice to simulate the voltage at the port. This is shown in the supplement. Then in MATLAB, we took the Fourier transform of this output voltage and multiplied it with the voltage-to-field transfer function at the point indicated by the red x in Figure \ref{fig:2}a to get the frequency response of the field in the load. Finally, we took the inverse Fourier transform of the field response, resulting in the time-domain plot shown in Figure \ref{fig:2}d. This process is shown in Figure S8 of the supplement. Here, in Figure 2d, we can see a steady decrease of the load ringing as we increase the damping resistance from \SI{1}{\ohm} to \SI{50}{\ohm}. At a damping of \SI{25}{\ohm} we see the fastest ringing-free transition, which corresponds to a 200 V transition in less than 1 ns. The underlying switching performance and its dependence on $\mathrm{R_{damp}}$ is consistent with the measurement on a similarly resonant load, the oscilloscope-GaNFET system resonant at $\sim$1.1~GHz, as shown by the ringing in Figure 1d.

We next worked to simplify the treatment of this load and remove ringing of the field at the point of interest. Generally, it is assumed that the wavelength of the exciting voltage is significantly larger than the length scale of the load, we can treat that system as a lumped circuit with parasitic quantities C and L associated with it \cite{pozar2009microwave}. Explicitly, we require

\begin{equation}
    \lambda = {v_\mathrm{{eff}}/f} >> \ell,
\end{equation} \label{eqn:1}

where $\lambda$ is the wavelength, $v_\mathrm{{eff}}$ the propagation velocity in the medium (in this case just $c$ for vacuum), and $f$ the frequency of interest. This approximation is generally held valid if the structure length-scale $\lambda$ is $\gtrapprox 10\ell$. 

We can see this lumped element behavior emerge explicitly in the impedance plotted in Figure 2c. At frequencies below 500 MHz, the parallel plates exhibit the characteristic of a 7.38 pF capacitor. As the frequency increases beyond 500 MHz and approaches the first resonance at 1.4 GHz, the load is no longer purely capacitive, but is well-modeled using an RLC resonant circuit with an inductance of 1.70 nH as shown in the Figure 2c. However, above 2 GHz, the system can no longer be treated as an 2nd-order LC lumped circuit due to the many resonant modes developed by the length scale of the load, some examples of which are shown in Figure 2b.   


This simplified RLC lumped element treatment offers significant advantage in the ability to estimate and remove ringing in our system. By treating the load as a lumped RLC model, it becomes significantly easier to understand the effects of the parasitic output capacitance and inductance of the driving FET. This treatment also allows “critical dampening” of the circuit with an appropriately selected series resistance, dramatically simplifying the filter design that would normally be needed for such a precise transition.  This is demonstrated in Figure 2d, where a damping resistance is selected to cause a highly smooth transition that resembles a critical damping in RLC circuits. Ringing-like behavior at $\sim$4 GHz and above still  does exist in the output, though it is significantly attenuated to below 1\% the transition magnitude. 

\begin{figure*} [t]
    \centering
    \includegraphics[width=\textwidth]{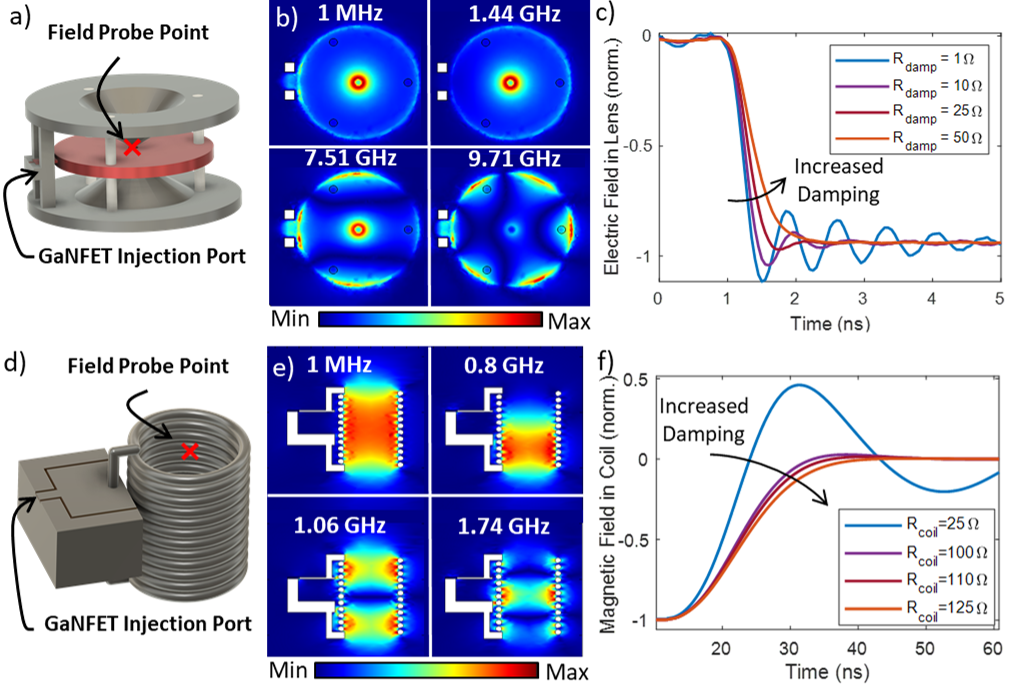}
    \caption{Other simulated loads. (a) A high voltage electron lens structure, with the red x showing the field sampling point near the center of the mirror. (b) The response of the lens at various frequencies, with resonances fitting the geometry of the structure. (c) The temporal field response with increasing damping. (d) An inductive load with the vertical magnetic field sampled in the center of the structure. (e) The magnetic field distribution, which is complicated even at low frequencies, and concentrates the field via capacitive coupling near the edge of the coils at higher frequencies. (f) The result of damping the inductor, with a huge ringing period due to the large DC inductance of the structure.}
    \label{fig:3}
\end{figure*}

We also examined other loads to validate the generality of this approach. In Figure 3a-c we show driving of a high voltage Einzel lens. This kind of structure is used commonly in electron optics, where the varied spacing of plates is designed to allow for high-strength fields near the center, while the wider spaced portion near the edges prevent electrical breakdown at the insulators during high voltage operation \cite{2017HandbookOptics}. This load has varied capacitance from the outside of the structure to the inside, as well as a much more significant inductive component due to the large ground return loop. Various resonances of this structure are shown in Figure 3b, and we can clearly see stronger electric fields near the center due to the closer plate spacing. Unlike the first capacitor model, this load cannot be easily simplified into a transmission line segment and thus matched to \SI{50}{\ohm} systems. However, the load can still be simplified into 5.50 pF capacitive model at low frequencies and a series RLC circuit with 2.35 nH inductance at higher frequencies. This fit is shown in Figure S9a of the supplement. This higher inductance and lower capacitance is consistent with the larger ground return path of this structure compared to the primarily capacitive load. As before, using GaNFET driving and dampening of the first RLC resonance, it was possible to generate a ringing-free field transitions in the load at the field probe point, this time in 1.2~ns, slightly slower. The generated field strength is equivalent to that that would be generated from a 200 V potential applied to the center electrode. It is interesting to note that nominally the first resonance of this lens and the first deflector are roughly the same, but this lens has a slightly slower response. This is due to the output capacitance of the GaNFET in the LTSpice model, which results in a response that is much more sensitive to the inductance of the load. This is discussed in the supplement in greater detail.

Another load we drove is the inductive coil shown in Figure 3a, which primarily generates a magnetic field rather than an electric field and is used commonly for applying strong magnetic fields to samples and for electron beam deflection systems \cite{2017HandbookOptics}. The driving topology for this circuit is different from that of the previous cases due to the low DC impedance of the load. This is shown in the supplement. At low frequencies, this element behaves as a near perfect 313 nH inductor, with an impedance that increases linearly as a function of frequency. As the frequency increases, the capacitance between the wires begin to shunt current, resulting in a reduced impedance and more concentrated fields near the edges as shown in Figure 3e. When the first resonance is reached around 260 MHz, we can model the circuit as a parallel RLC circuit with a capacitance of 1.15 pF. Similar to the previously discussed loads, we were able to tune $\mathrm{R_{coil}}$ to achieve a ringing-free magnetic field response driven by a \SI{20}{\ampere} current as shown in  Figure 3f. However, due to the large inductive component coupled to the GaNFET's output capacitance, this load requires a larger resistance of 110 $\Omega$ to damp, slowing the response substantially. Although the current transition time of roughly \SI{20}{\nano\second} can be shortened by reducing the loop size or number, inductive loads will almost always be slower that capacitive loads due to the fixed output capacitance of the GaNFET circuit and layout.


\section{Discussion}
Besides providing fast, ringing-free switching of hundreds of volts and tens of amperes, the GaNFET switches described above also offer the following useful features. (1) Affordability. The circuit shown in Figure 1b is simple to assemble using standard soldering tools, and costs two-orders-of magnitude less than commercial pulsers. This enables wide use in a range of experimental apparatus. (2) Compatibility with computer and FPGA driving. These circuits also offer simple and arbitrary control by 3.3 V pulses on \SI{50}{\ohm} lines; (3) repetition rates up to 60 MHz (if a high side P-Channel transistor is used rather than the resistor), in comparison to several KHz for spark gap technologies \cite{Rahaman2010ARate} and 25 MHz for custom state-of-the-art avalanche circuits \cite{Beev2017Note:Rate}; (4) 2.5 ns propagation delays which open the possibility of real-time control based on triggering of other single-shot measurements in an apparatus; (5) a <\SI{0.25}{\square\centi\meter} active form factor that allows for the integration of these circuits extremely close to the loads they drive, minimizing parasitic energy storage that slow transitions; (6) <1 W power dissipation at 10 MHz, which scales approximately linearly with the circuit repetition rate. The low power dissipation means that it is possible to operate these circuits under vacuum, a requirement for driving loads such as charged particle optics in this manner.

The pulsers discussed so far have been optimized for single-sided driving, where one transition (the negative sloped one) is on the order of nanoseconds, while the reset transition is on the order of tens of nanoseconds. This maximizes the speed of the faster edge and simplifies the circuit design significantly, though limits the technique in terms of repetition rates and for some applications. In order to drive both positive and negative edges, alternative topologies such as a half-bridge configuration with bootstrapping is required. Commercial drivers, such as the LMG1210 from Texas Instruments, do exist for this, though the increased size and complexity of the circuitry slows their transition times. Also note that fully integrated GaNFET circuits do exist, such as the LMG3410 from Texas Instruments, though have slower switching transitions than demonstrated here.

By varying the FET technology and topology, this approach can be extended to switching even higher voltages and currents. The simplest way of doing this is to vary the GaNFET technology, using for example the GS61004B from GaN Systems or the TP65H070LSG from Transphorm, both of which are capable of 650 V operation, though their increased gate capacitance, output capacitance, and parasitic inductance lead to slowed output transitions. Another approach would be to use SiC technology, allowing for transitions of several thousand volts though at the cost of even slower transitions. Alternate topologies used historically to improve silicon performance, such as parallel and stacked FETs \cite{Jiang2007FastMOSFETs} or cascode configurations \cite{Baker1992AMOSFETs} could also be used to improve both voltage ratings and transition performance in these GaN systems.

Several routes also could be used to increase the speed of these switches to achieve subnanosecond transitions. Smaller, lower voltage FETs such as the 100 V EPC2037 from the Efficient Power Corporation would have significantly faster switching due to its smaller gate and output capacitance, as well as its lower form factor. For low voltages and medium currents of $\sim$5 A, GaNFET gate drivers such as the LMG1020 itself could also be used directly to drive loads with $\sim$400 ps transition times. If state-of-the-art rise times are desired, non-linear pulse sharpening techniques such as step recovery diodes \cite{WongChoi2011Note:Circuit} and nanoplasmas \cite{SamizadehNikoo2020Nanoplasma-enabledElectronics} could be used together with these FETs, which would allow for transitions on the order $\sim$100 ps or less, and would not increase layout size significantly. 

\begin{figure}
    \centering
    \includegraphics[width=0.48\textwidth]{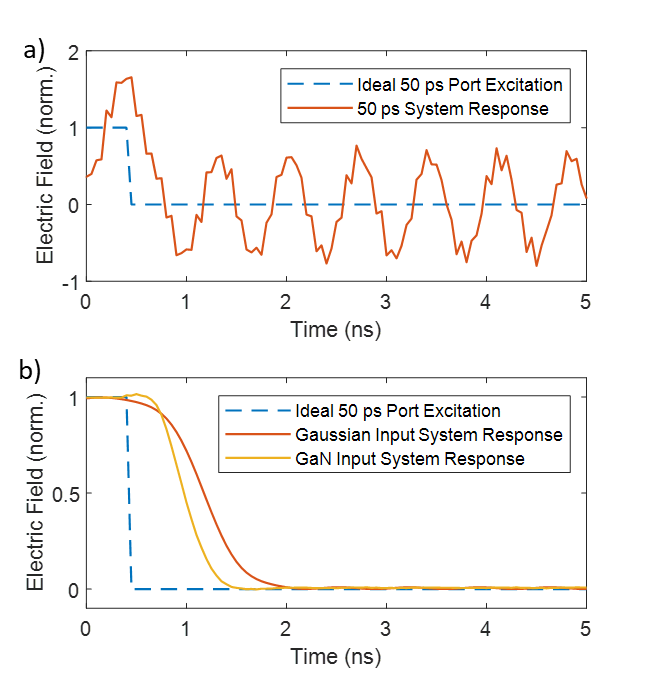}
    \caption{Step response of the load in Figure 2a. (a) The response of the load to a 50 ps step (dotted line), set by the bandwidth of the simulation. This shows ringing at not just the lowest resonant frequency, but also at the higher order frequencies shown by the roughness on the trace. (b) The same response, but with the edge slowed by a Gaussian filter to reduce the ringing to 1\% the transition magnitude. The orange trace is taken from Figure 2d for comparison, showing an even faster RLC-based response.}
    \label{fig:4}
\end{figure}

A natural question that arises as we increase the speed of the transitions is how fast the loads fundamentally can be driven, especially when trying to ensure no ringing. Even if we drove the load with an ideal 100 ps step, the high-frequency resonances of the load would drown out the signal in ringing and drag out the transition. This is demonstrated in Figure 4a for the real load in Figure 2a. If we then put this transition through an ideal Gaussian filter to reduce the pulse bandwidth and reduce the ringing to $\sim$1\% the transition, we get the result in Figure 4b which is significantly slower, taking roughly 2 ns to get within 1\% accuracy. This transition is significantly slower than the RLC response also shown in Figure 4b, which took only 1.2 ns to achieve equivalent accuracy. Different filtering schemes, such as Chebyshev or Butterworth filters might improve this further, but these were not explored.

Two general approaches come to mind for driving these loads faster: optimizing the load structure to minimize resonances, and shaping the incident signal to avoid these resonances. The first relies on pushing the load resonances up or reducing their magnitude. Increasing the resonances can be achieved by reducing the physical size of the load and avoiding the use high permeability and permittivity materials. Reducing their quality factor can be done by introducing surface roughness to the load, increasing the structures resistivity with different materials or coatings, or introducing a lossy propagating media. However, we have not explored any of these routes in depth.

The input engineering approach can be realized with various schemes of increasing complexity. The simplest approach would be to use a low-pass filter to generate a transition with a sharper cutoff that excites the first resonance less than the Gaussian filter we simulated. Exactly how much faster you could drive this circuit with that is unclear and depends strongly on the load and specific filter used. A second approach that would give a faster response would be to design a resonant filter that selectively removes the strongest resonant peaks in the output. Exactly how accurately one could find and remove these resonances and to what bandwidth is unclear. A third approach similar to the resonant one would be to use an arbitrary waveform generator (AWG) to digitally generate create an input signal with the resonant modes removed. This would remove ringing to the bandwidth and gain to the limits of the AWG and amplifier used, though again probing to precisely remove the ringing would be a challenge. This approach has been explored in recent work \cite{Klopfer2020RFMirrors}.


\section{Conclusion}
In this work, we demonstrated the potential of inexpensive GaN power electronics to switch real loads commonly found in physics and engineering with high power and few-nanosecond, ringing-free transitions. We first validated this concept by constructing a test-bed circuit capable of switching \SI{100}{\volt} in 250~ps with ringing, and 1.5 ns without ringing. This was done with a design using raw parts costing two orders of magnitude less than comparable commercial pulse generators. Switching up to 200 V was tested to ensure the switch did not fail, but was never measured in operation due to concerns of damaging the oscilloscope used. Due to our inability to reduce measurement parasitics at such high voltages, and in order to see the field distribution with a real load, we simulated the output of the circuit for various structures ranging from capacitive to inductive. We then extracted the temporal response of the electric and magnetic fields in the load at various points, demonstrating $\sim$1 ns, ringing-free switching for the provided capacitive loads. We also demonstrated a model of each load's lowest resonance as an RLC resonator, which allowed us to simplify analysis of the load and damp the dominant ringing of the field. We demonstrated that although higher frequency components theoretically remained, they were negligible and the switching behaved much like a critically-damped circuit. 

We noted that this pulser is theoretically capable of being controlled directly and arbitrarily by 5 V logic signals up to 60 MHz, with $\sim$2.5 ns propagation delays, limited in the current iteration by circuit resetting from the biasing resistance. We also discussed the the possibility of integrating this circuit in environments such as vacuum as well as using other pulsing topologies to increase the current and voltage ratings even further. We finally discussed the limits of real load switching and demonstrated this work provides a faster response than ideal Gaussian low-pass filtering can provide. Future work will expand upon these results by integrating these circuits under vacuum with deflectors and measuring the time dependence of GaN-switched loads directly using electron beams, the smallest probe imaginable. We believe this work will have significant application throughout physics where inexpensive, simple, high voltage, and precise switching is needed.

\begin{acknowledgments}
This work was supported by the Gordon and Betty Moore Foundation. This material is based upon work supported by the National Science Foundation Graduate Research Fellowship under Grant No. 1745302. Yugu Yang-Keathley and Ben Slayton acknowledge support from the Douglas D. Schumann Professorship. The authors thank Marco Turchetti and Navid Abedzadeh for help with designing the original structures simulated, Texas Instruments for providing internal SPICE models of the LMG1020 used in the simulations, and the QEM-II collaboration for insightful discussions. In particular, we would like to thank the Kasevich group at Stanford, especially Adam Bowman, Brannon Klopfer, and Stewart Koppell, for many discussions of fast pulsing technology, alternative techniques to drive such loads, and applications of these pulsers. The authors also thank Ilya Charaev and Owen Medeiros for helpful feedback on the manuscript. 

\end{acknowledgments}

\bibliography{references}

\end{document}


\author{John W. Simonaitis}
 \email{johnsimo@mit.edu}
 \affiliation{Research Laboratory of Electronics, Massachusetts Institute of Technology
}%
\author{Benjamin Slayton}%
\affiliation{%
Wentworth Institute of Technology
}%

\author{Yugu Yang-Keathley}
 \affiliation{%
 Wentworth Institute of Technology
}%

\author{Phillip D. Keathley}
 \affiliation{Research Laboratory of Electronics, Massachusetts Institute of Technology
}%

\author{Karl K. Berggren}
 \affiliation{Research Laboratory of Electronics, Massachusetts Institute of Technology
}%

\date{\today}

\title{Precise, Subnanosecond, and High-Voltage Switching of Complex Loads Enabled by Gallium Nitride Electronics: Supplementary Information}
\maketitle

\renewcommand{\thefigure}{S\arabic{figure}}
\renewcommand{\thetable}{S\arabic{table}}
\renewcommand{\theequation}{S\arabic{equation}}

\section{Circuit Design}
\begin{figure}
    \centering
    \includegraphics[width = 0.9\textwidth]{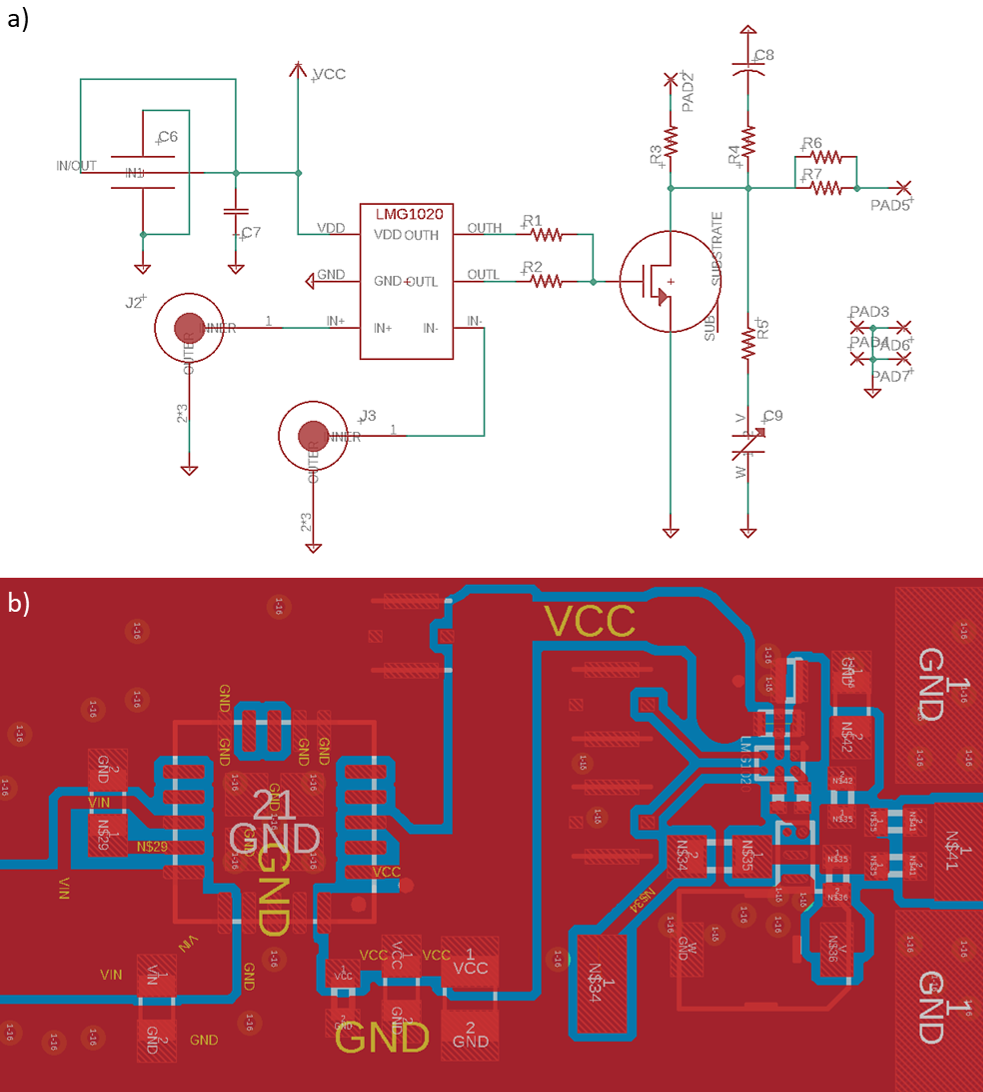}
    \caption{Test-bed circuit schematics and layout. (a) The main part of the circuit design, excluding the power supply which is represented by VCC. (b) The layout. The active region is to the right. Note the proximity of the input capacitors to the LMG1020 and the minimized ground return path directly under the drive loop.}
    \label{fig:circuitLayout}
\end{figure}

The full circuit used is shown in Figure \ref{fig:circuitLayout}a, with part numbers labeled. The power supply is based on the TPS7A4700RGWT from Texas Instruments, though any low-noise DC supply will suffice. In order to minimize switching times and ringing, significant care was taken to choose driver capacitors with high self resonant frequencies (SRFs) and low equivalent series resistances (ESRs). We placed the capacitors as close to the LMG1020 driver as possible, as shown in Figure 1b. We used the combination of a \SI{0.1}{\micro\farad} 0402 feed-through capacitor (TDK YFF15PC0J474MT) closest to the driver, and a \SI{1}{\micro\farad} 0204 (wide package) capacitor (TDK C0510X5R0J105M030BC) slightly further out to provide a low impedance drive loop and maintain biasing to the driver respectively. The temperature stability of the capacitors was sacrificed in order to minimize the package sizes used, with X5R ratings chosen rather than X7R. Low heat dissipation in our system and good thermal grounding ensured this did not cause problems in our setup. The most relevant parts and part numbers used are listed in table \ref{tab:components}.
\begin{table}[b]
    \centering
     \begin{tabular}{c  c  c  c} 
     \hline
     Label(s) & Value  & Manufacturer & Part Number \\ [0.5ex] 
     \hline
     C6 & \SI{0.1}{\micro\farad} & Murata Electronics & NFM15PC104D0J3D \\ 
     C7 & \SI{1}{\micro\farad}  & TDK Corporation & C0510X5R0J105M030BC \\
     C8 & \SI{20}{\pico\farad}  & Vishay Vitramon & VJ0603D200JXPAJ \\
     C9 & 8 - \SI{40}{\pico\farad}  & Knowles Voltronics & JZ400HV \\
     R1, R2 & \SI{2}{\ohm}  & Vishay Dale & CRCW02012R00FXED \\
     R3 & \SI{1}{\kilo\ohm} & Bourns Inc. & CR0603-JW-102ELF\\
     R4, R5 & \SI{10}{\ohm} & Vishay Thin Film & FC0402E10R0BTT0\\
     R6, R7 & \SI{25}{\ohm} & Vishay Thin Film & FC0402E25R0BTT0  \\
     J2, J5 & \SI{12}{\giga\hertz} & Murata Electronics & MM5829-2700RJ4 \\
    \end{tabular}
    \caption{List of components from schematic shown in Figure \ref{tab:components}}
    \label{tab:components}
\end{table}

We also sought to minimize the size of the drive ground loop. This was done by the use of a thin two-sided \SI{100}{\micro\meter} Kapton substrate (PCBWay) with a direct ground return trace running under from the GaNFET source pad directly to the driver ground as shown in Figure \ref{fig:circuitLayout}b. The bottom solder mask was removed, and the circuit directly mounted to a copper plate for heat sinking under vacuum.

Damping was implemented using high-frequency thin film resistors ranging from \SI{10}{\ohm} to \SI{200}{\ohm} (Vishay Thin Film FC Series). Though the 0.125 W power rating would indicate they can only operate to 1.25 MHz (assuming 100 V switching of 10~pF), we found the resistor continued working to ~5 MHz. Using better heat-sinking and multiple damping resistors in parallel to spread out the dissipation, we found this circuit could operate up to 20 MHz, the maximum repetition rate we tested. However, in this condition, the load never fully recharged due to the large bias path impedance. This exercise was primarily aimed at measuring power dissipation limits. If a faster resetting rate is needed a high-side P-channel FET would need to be put in place of the damping resistor, or a bootstrapping circuit based on the LMG1210 used.

We were also able to make these circuits vacuum compatible. This was done by using silver-based solder (SMD291SNL-ND from 	Chip Quik Inc.) and low-outgassing Kapton substrates (PCBWay). After assembly, the circuits were mounted on Oxygen-free high  conductivity (OFHC) copper, sonicated in PCB cleaning solution (PELCO Kleensonic™ APC) rinsed by water and IPA, and finally encased in thermally conductive and electrically insulating ceramic epoxy (EPO-TEK® H70E Thermally Conductive Epoxy).

 \section{Assembly Procedure}
The following details the construction of this high-speed GaN circuit.

\begin{enumerate}
  \item First, we prepare our work space. Electrostatic discharge (ESD) safe mats and grounding bracelets are a necessity to get high circuit yields. We first preheat a hot plate to \SI{190}{\degree}C. We then preheat a soldering iron to \SI{300}{\degree}C. Tape the circuit using an easy release masking tape to whatever work surface you hope to assemble it on, as shown in Figure \ref{fig:assembly}a.
  
  \item Next, we use a soldering stencil from the layout given (also from PCBway) and align it using a magnifying glass or microscope to the pattern. Tape one side of the mask as shown with the flipped stencil in Figure \ref{fig:assembly}a. We found aligning to the smallest component pads was the easiest way to do this.
  
   \begin{figure} [b]
    \centering
    \includegraphics[width=0.9\textwidth]{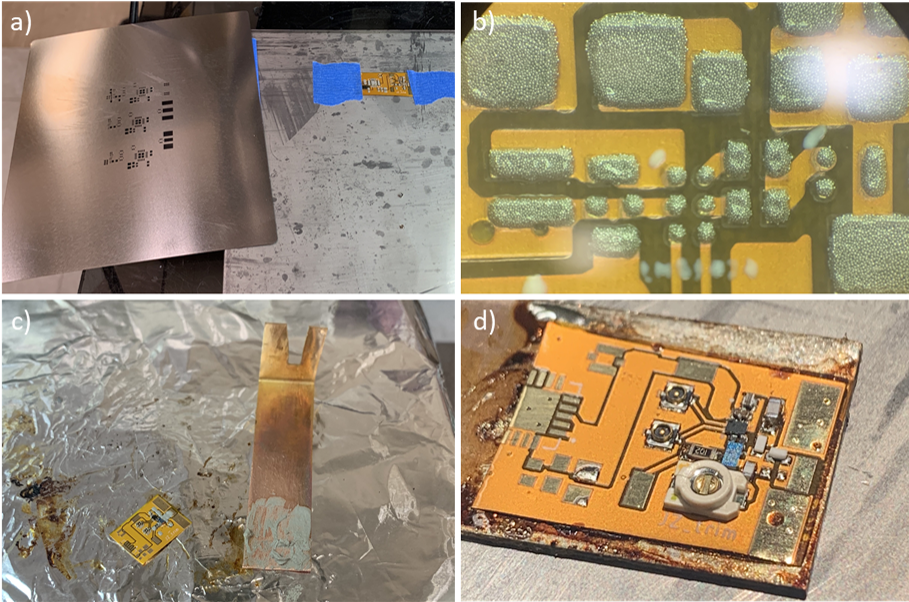}
    \caption[Assembly of Circuit]{Overview of circuit assembly for thermal management and vacuum compatibility. (a) Taping of the circuit to the board, and the stencil attached to the left with tape (flipped upside-down). (b) Solder paste clearly on pads, resulting from a well-aligned stencil. (c) Heating of the circuit on a hot plate to solder each connection. Also shown is the solder spread on the copper mount, which we place the circuit on for heat-sinking. (d) the final circuit without wires, showing successful soldering of the components and bonding to the copper heat sink.}
    \label{fig:assembly}
    \end{figure}

  \item Apply a small amount of solder paste to the edge of the mask. Holding the stencil firmly and evenly down, use a straight edge (such as a plastic card) to gently spread the paste over all of the holes, ensuring each hole is covered. Then, pressing more strongly on the card, scrape away any excess material.
  \item Remove the stencil, being careful to pop it directly up.  Inspect the layout, ensuring there are no solder paste connections between different pads. If there are only a few, it is possible to correct by breaking the bridges with a fine tip such as some tweezers. If there are too many bridges or the paste is too spread out, wipe the surface clean and repeat. An example of good looking paste application is shown in Figure \ref{fig:assembly}b.
  \item Now assemble components. If not already done, make sure to wear an ESD bracelet, or at least ground your tweezers with an alligator clip or something similar. Even the slightest discharge will destroy the GaNFET, causing your FET to be measured as approximate \SI{1}{\ohm} to \SI{10}{\ohm} across the source to drain even when off. Other various tips are below.

  \begin{itemize}
        \item When assembling, we generally we start with the largest components and work down in size, finishing with the EPC2012c and LMG1020 as their pads are the most delicate and they are the most expensive. The only exception to this size rule is placing R1 and R2, which we generally do fairly early since they are inexpensive and easy to mess up, and it is nice to have space to place them without worrying about knocking the GaNFETs or driver out of place. 
        \item Various methods exist for stabilizing hands while doing this such as resting your hand on your fingers, exhaling during the final component placement, and avoiding caffeinated beverages the day of assembling. Using a clean wipe and alcohol, clean the tweezers after each placement to prevent solder paste from sticking the component to the tweezers. 
        \item If the component is misplaced, put the tweezers tip onto the PCB to stabilize shaking, and then slide the tweezers tip around to nudge components into place. As long as the component pads make contact with the solder paste on each pad, it should be fine as surface tension will pull components exactly into place.
  \end{itemize}
  
  \item Place the circuit onto the hot plate. If available, place a fume extractor or fan above the hot plate. Increase the plate temperature to \SI{290}{\degree}C, though this can be adjusted depending on the plate, as long as the PCB does not darken the temperature is not too high. As this happens, you should see the solder begin to melt. Watch the smallest components, especially the LMG1020 and EPC2012c, and make sure the components get pulled into place by surface tension of the solder. If this does not happen, try slightly nudging the components with tweezers to knock them into place. If when you touch the components, there is no action trying to return them to place, they are likely not properly placed or there is not enough solder.
  
   \item On the piece of OFHC copper you wish to mount your circuit on, spread out the silver solder paste evenly. For our thin heat sink this is shown in Figure \ref{fig:assembly}c. Remove the assembled circuit and place it onto the copper. The solder should harden holding the circuit components in place. Now, place the entire copper piece onto the hotplate. Wait until the solder melts and the circuit board is pulled into place by surface tension.
   \item Reduce the hot plate temperature to \SI{190}{\degree}C. Wait until all of solder re-solidifies. Once this happens, take the soldering iron and apply it to the biasing and power pads and solder on connection wires. Remove the circuit. Clean the circuit with acetone and IPA. If vacuum performance is required, sonicate in PCB cleaning solution given in the materials section. If high repetition rate performance is required, encase in EPO-TEK H70E thermally conductive epoxy (PELCO). The end product should look like Figure \ref{fig:assembly}d.
   \end{enumerate}
   
\newpage
   
\section{Circuit Safety and Testing}
    Next, we test the circuit. We first connect the circuit to ground. Next, we attach the high voltage biasing port to a high voltage port. In our case, we used a 150~V piezo supply (Thorlabs model MDT693B) with a \SI{1}{\mega\ohm} resistor tied to ground (for safety and ESD protection) coupled through a BNC cable (rated to 300~V). This high voltage supply is left off until testing. We then turn on the power supply supply. In early tests we used a 6~V power supply (XP Power VEL05US060-US-JA) with the designed regulator circuit. For later tests, to reduce power dissipation in the circuit and it more vacuum compatible, we used a 5~V direct connection from a low noise power supply (Keithley Model 2220). We saw no difference in results using either source.

      \begin{figure}[b]
        \centering
        \includegraphics[width=0.9\textwidth]{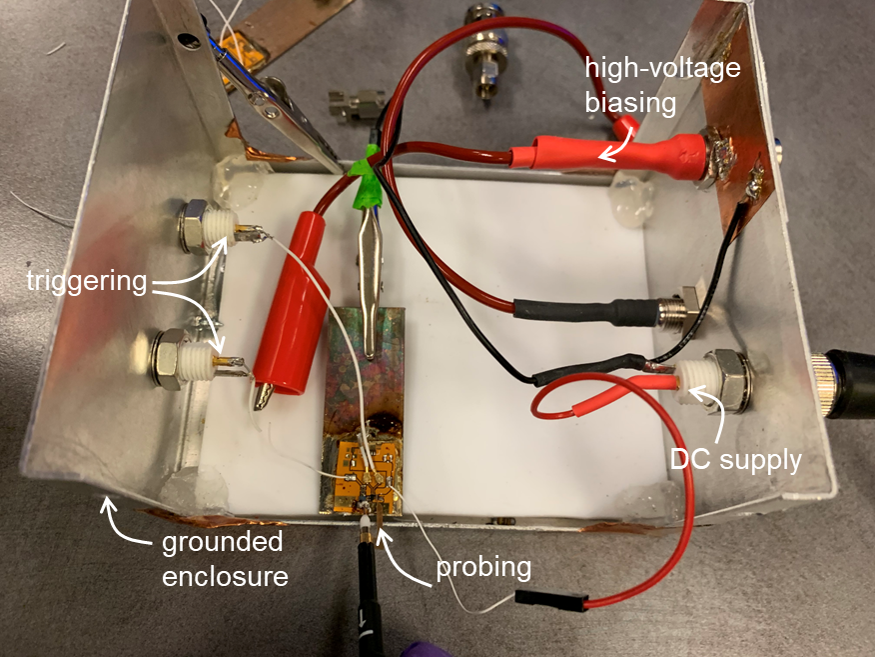}
        \caption[Testing Setup]{Testing setup used, showing shielding, ground insulation, and probing.}
        \label{fig:testingEnclosure}
    \end{figure}
    
    We then hook up the RF ports. We used an SMA-to-JSC adapter (GradConn CABLE 366 RF-200-A) and BNC-to-SMA adapter to get the input signal from the signal generator (Keysight 33250A) to the driving board. For traces shown, we input a 3~V, 100~ns wide pulse with 5~ns rising edges at a repetition rate of 100~KHz. This low rate was chosen to reduce the voltage offset caused by AC-coupling, reduce thermal effects, and ensure we did not put too much power into the oscilloscope. Using the probe, which protects the oscilloscope, we tested to 20~MHz, without failure, though this limited by the recharge time of the biasing resistor which reduced the full voltage swing of the circuit. The second input was grounded.
   
    We then turned on the high voltage supply, starting at 5~V to ensure the circuit functioned properly. Once this was verified we slowly increased to voltage to the set point. For initial tests we used a LeCroy PP007-WR 500~MHz \SI{10}{\mega\ohm} probe, tested up to 100~V with a blade ground connector, as shown in Figure \ref{fig:testingEnclosure}. The probe was compensated using a 1~MHz square wave signal from the signal generator.

\begin{figure}
        \centering
        \includegraphics[width=\textwidth]{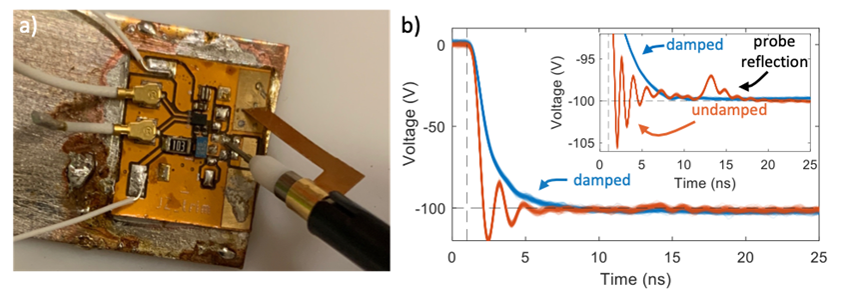}
        \caption{Passive probing. (a) Close up on the grounding connection used. (b) Damped and undamped traces, showing recurring ringing we hypothesize are due to probe reflections.}
        \label{fig:passProbe}
    \end{figure}

    A zoom in of this testing, as well as the resulting traces is shown in Figure \ref{fig:passProbe}a and b respectively. Both traces were acquired with 100~V biasing. The undamped trace in Figure \ref{fig:passProbe}b shows significant ringing, and a recurring ringing at 12~ns. We hypothesis that the recurring ringing around 12~ns is due to reflections in the probe line, as their occurrence matches the round trip time of the line (knowing the cable length is 1.2~m, and assuming the wave propagates at ~66\% the speed of light, a common value for many coaxial cables). The damped trace shows a well-damped transition. However, as the probe reflection indicates, this system can not be easily modelled as an RLC-lumped element circuit since the cable length scale is similar to the transition time. 

\begin{figure}
        \centering
        \includegraphics[width=0.7\textwidth]{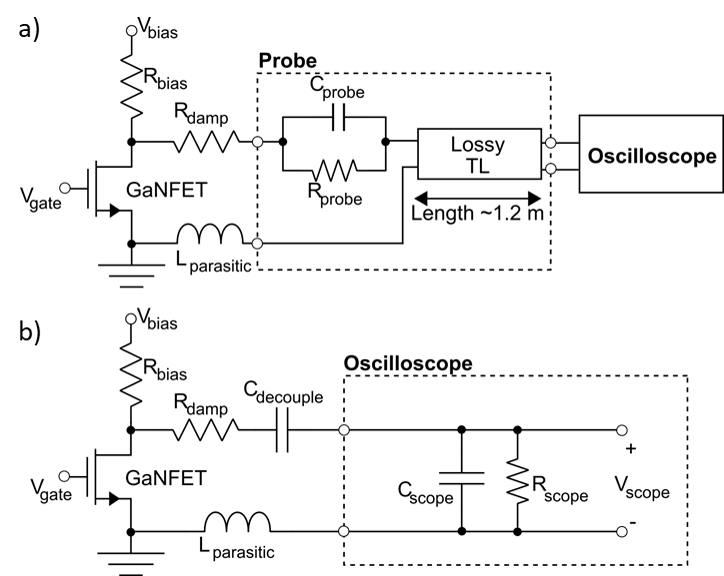}
        \caption{Comparison of measurement types. (a) Passive probe model. (b) Direct connection model.}
        \label{fig:measComparison}
    \end{figure}

    For the final testing, we directly hooked up our circuit through an SMA connector which we adapted to a BNC connector and inserted into the oscilloscope. This is described in the main text. In Figure \ref{fig:measComparison}a, we show models of each measurement. Figure \ref{fig:measComparison}a shows our model for the passive probe. This consists of a capacitance $C_{probe}$ in series with a lossy transmission line (TL) with a length of ~1.2~m. This length scale results in a roughly 100~MHz resonance in the load we drive. Though this resonance is designed to be weak by making the cable highly lossy, this length scale still breaks down our analysis of this system as an RLC-type load. The observed reflection in \ref{fig:passProbe}b is evidence of this. For simplicity, compensating circuitry in the probe is not shown.
    
    \begin{figure}
        \centering
        \includegraphics[width=0.8\textwidth]{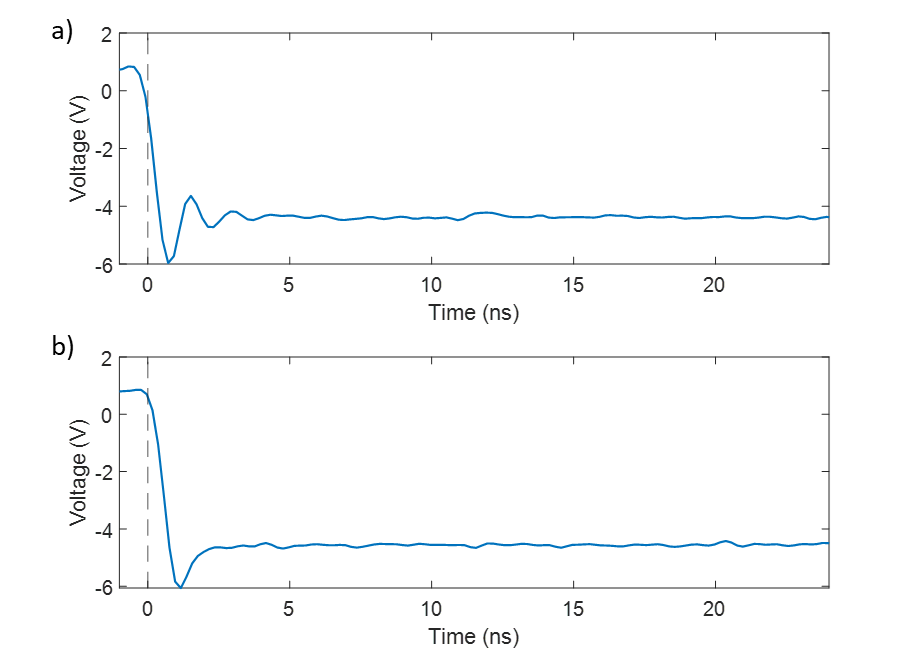}
        \caption{Effect of changing gain on the measurement. (a) A trace with a gain of 1.02~V/division. (b) The same trace with a gain of 1.00~V/division. Switching this range causes an audible click in the oscilloscope, likely due to some sort of switched amplifier configuration on the input, which visibly changes the response of the circuit.}
        \label{fig:changingTrace}
    \end{figure}
    
    Figure \ref{fig:measComparison}b shows the resulting model from directly hooking up our circuit to the oscilloscope. This simply results in $C_{decouple}$ being placed in series with $R_{damp}$ and $L_{parasitic}$, creating a simple RLC-type resonator that we can damp. By carefully selecting the value of $C_{decouple}$, we can then attenuate our voltage signal, allowing us to safely probe voltages up to 100~V. If this capacitor is not used, changing the range of the oscilloscope changes the measured ringing of the circuit. This is demonstrated in Figure \ref{fig:changingTrace}. We hypothesize that this is due to the oscilloscope amplifier changing impedance as the gain changes. By decoupling the impedance with a much smaller capacitor, we then stabilize the impedance seen by the GaNFET, and thus the measurement.

\section{Simulation details}
\begin{figure}
    \centering
    \includegraphics[width = 0.97\textwidth]{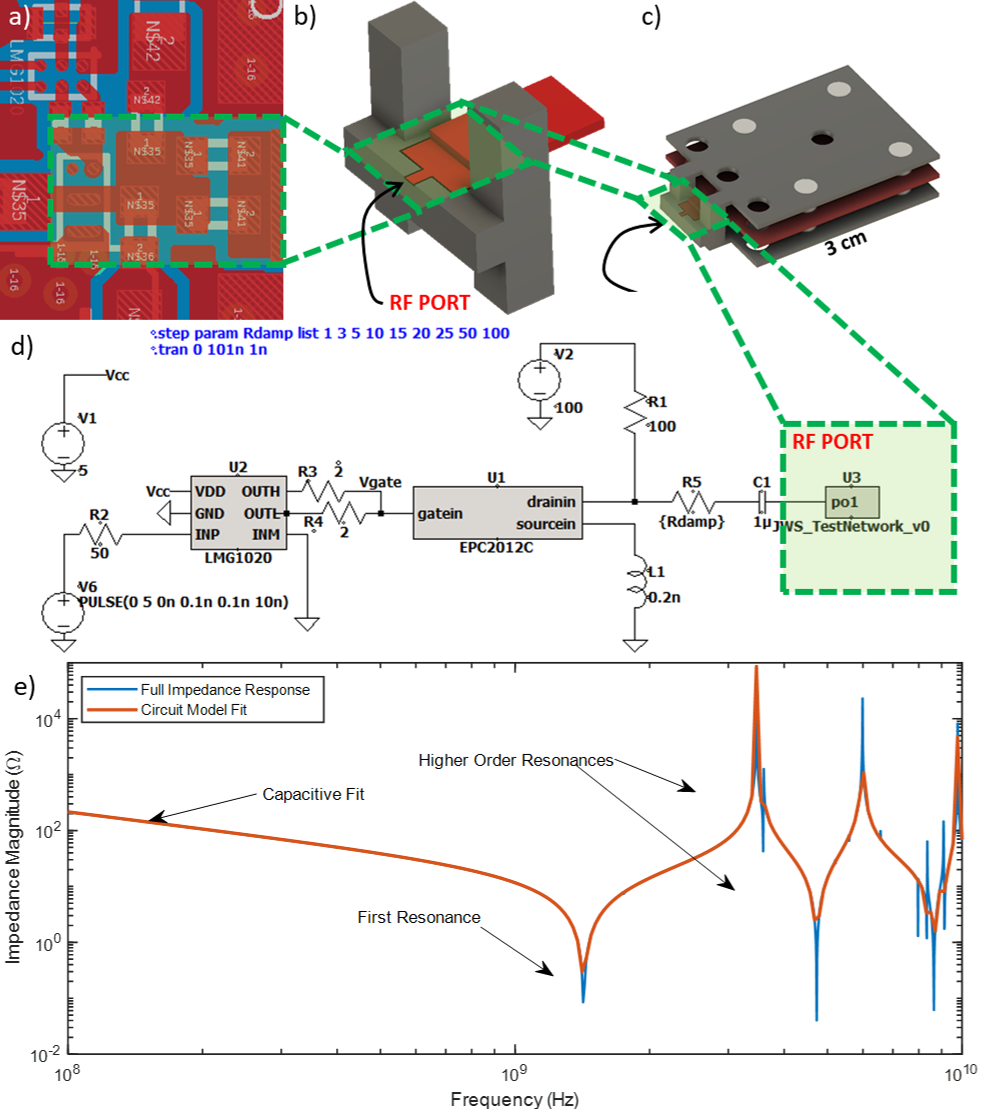}
    \caption{Representation of the simulation. (a) A close up of the PCB, with the area treated in the simulation highlighted in green. (b) The simplified PCB model that was used to launch the RF signals in all of the simulations. (c) The RF port attached to the load of interest, in this case the deflector. (d) The SPICE simulation, which interacts the EPC2012C and LMG1020 with the RF port, denoted as JWS\_TestNetwork\_v0. (e) A comparison of the full impedance response from COMSOL to the circuit model fit generated in MATLAB from the rationalfit and generateSPICE functions, and output as JWS\_TestNetwork\_v0. Note that the rational fit model smooths some features of the COMSOL simulation, shown by the deviations of the blue and orange curves around 3.5 GHz and the three blips from 7 - 9 GHz shown. These individual points are likely not physical and due to glitches in the simulation.}
    \label{fig:PCB_Port}
\end{figure}

The first modeling step of this work was to represent the printed circuit board (PCB) layout and the load in COMSOL. This was done by extracting the dimensions of the PCB and drain pad of the GaNFET from the layout to Autodesk Fusion 360 and attaching this to various structures we designed. This PCB port model is shown in Figure \ref{fig:PCB_Port}a. The models of these structures were then imported into COMSOL.

The simulation consisted of a frequency-domain calculation of the loads using the RF Electromagnetics Module of COMSOL Multiphyics, exporting of the S-Parameters to MATLAB and using MATLAB's RF Toolbox rationalfit function to generate a rational fit to the model. The simulation was run from 10 MHz to 10 GHz at 10~MHz steps. Then using the MATLAB generateSPICE function we exported a SPICE circuit fitting this resonant behavior up to 10~GHz. A comparison of this fit is shown in Figure \ref{fig:PCB_Port}e.

The LTSpice circuit model used is shown in Figure \ref{fig:PCB_Port}d. Note that the circuit is AC-coupled by a \SI{1}{\micro\farad} capacitor. This capacitor was used because the rational fit model is unable to correctly represent the capacitor's infinite impedance at DC. Figure \ref{fig:PCB_Port}e shows the impedance response of this combined system, which has a plateau around 1 MHz. The \SI{1}{\micro\farad} decoupling capacitor allows the circuit model to treat the infinite DC impedance properly, and as long as this capacitance is much larger than the system under test, has virtually no effect on the results. The existence of this plateau should not effect the performance of this simulation, as the impedance here is already significantly higher than that of the rest of the circuit.

\begin{figure}
    \centering
    \includegraphics[width = 0.98\textwidth]{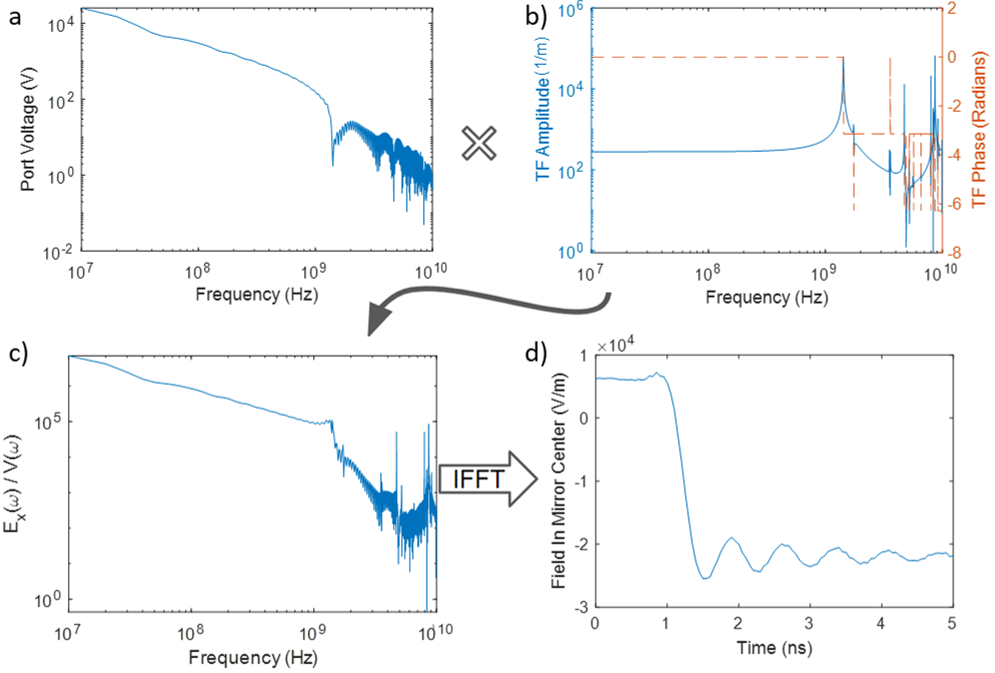}
    \caption{Fourier calculations of field ringing in the loads. In this example we show that multiplying (a) the frequency spectrum of the pulse by (b) the port voltage to field in load transfer function (the top right, taken at the red x in Figure 2a) results in the spectrum in (c). Note that (b) is basically flat until ~500 MHz, then begins to rise until a peak at ~1.44 GHz. This occurs because a quarter-wave mode that forms in the mirror with the minimum at the port, and maximum at the edge of the deflector as shown in Figure 2b at 1.44 GHz. This means that if we force a small set field (normalized to 1 V) at the input port near this resonance, a large field will result at the edge, and thus a large field at the sampling point as well. Depending on the distance of the sampling point away from the port, the frequency at which this peaks changes. The closer the sampling is to the port, the higher frequency this resonance. Taking the inverse Fourier transform results in the time domain response shown (d).}
    \label{fig:FT}
\end{figure}

Next, we took the time domain output from LTSpice, interpolated it to match the COMSOL frequency stepping, and took the Fourier transform of this. This resulted in the plot shown in Figure \ref{fig:FT}a. We then multiplied it with the voltage to field transfer function of the load (Figure \ref{fig:FT}b) resulting in the output seen in Figure \ref{fig:FT}c. Taking the inverse Fourier transform of this led to Figure 2d in the main text.

\section{Analysis of Other Loads}
\begin{figure} [h]
    \centering
    \includegraphics[width = 1\textwidth]{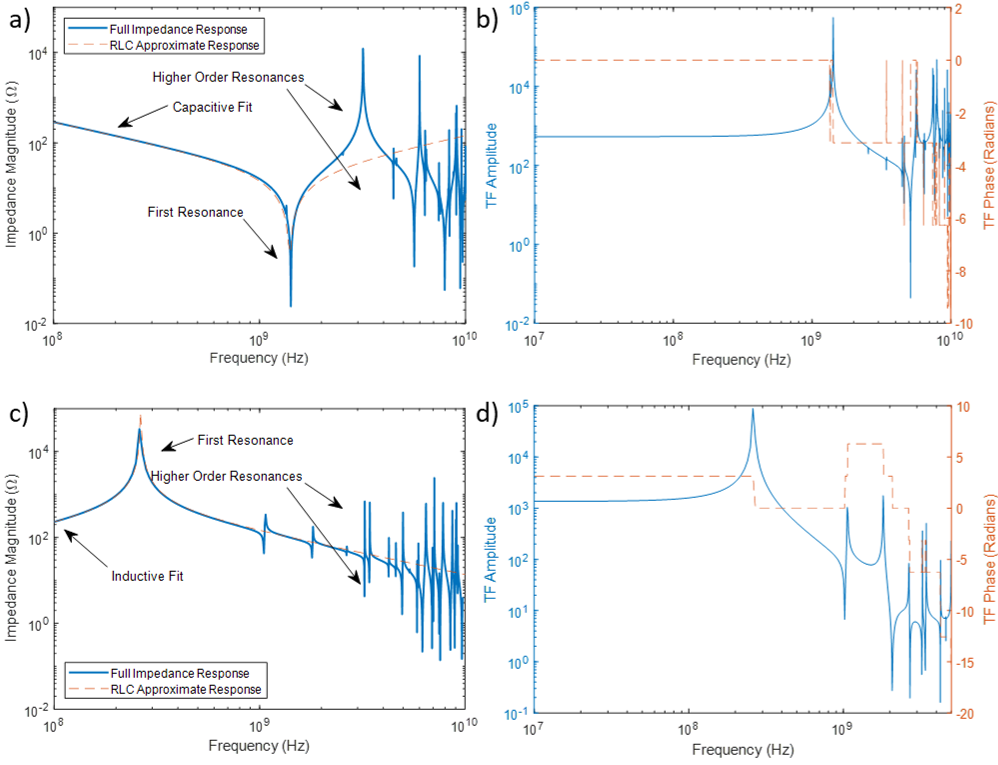}
    \caption{Example impedance plots and transfer functions of the other loads shown. (a) The port impedance of the Einzel lens, with the fit to a series capacitive-inductive load. (b) The transfer function relating the field in center of the lens to the voltage at the input port. (c) The port impedance of the inductive coil load. The resonance is much lower than in the other cases due to the large inductance of the coil. (d) The transfer function of the magnetic field in the center of the coil to the current input into the load.}
    \label{fig:altLoads}
\end{figure}

The impedance response and transfer functions of the Einzel lens and coil are shown in Figure \ref{fig:altLoads}a-d. From the RLC fits shown we extract the lumped parameters of the loads given in the main text. Note that for the inductor model we only simulated up to 5 GHz due to the lower resonances of the structure which lead to increasing complex and likely non-physical resonances in the 5-10~GHz range.

\begin{figure}
    \centering
    \includegraphics[width = \textwidth]{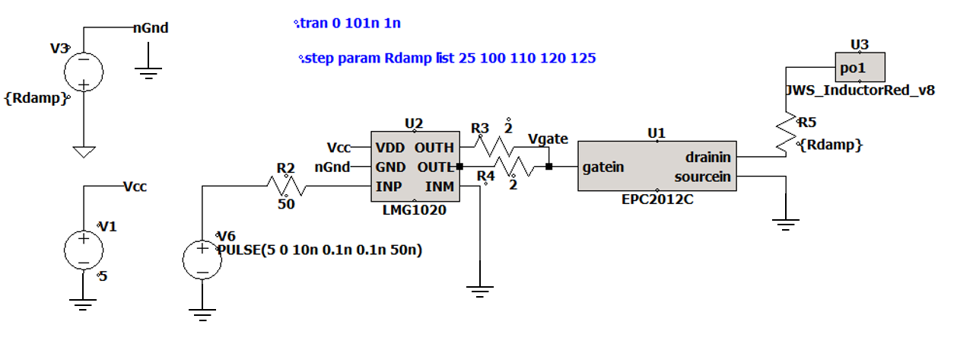}
    \caption{Simulated inductor topology. Notice the replaced ground symbol attached to nGnd. This was used because the circuit model of the inductor was internally referenced to ground, so in order to drive the inductor as if it was attached to a voltage source to the drain, we had to drop the potential of the rest of the circuit to this so-called "new" ground.}
    \label{fig:L_SpiceModel}
\end{figure}

The inductive load was unable to be driven effectively by the circuit topology in Figure 1a of the main text due to the significant impedance differences of inductive and capacitive loads. The SPICE model used is shown in Figure \ref{fig:L_SpiceModel}. The damping approach for the inductive load is not very efficient for real driving, but still demonstrates the fundamental switching capabilities of the GaNFET. Alternative damping techniques such as the use of a series resistor and capacitor in parallel would probably be more feasible, though was not explored. 

Order-of magnitude estimates can be made for these loads without the need for full simulations. For inductive loads, we can estimate the load inductance using equation \ref{eqn:L_ind}, where $A$ is the loop area, $\mu$ the core permeability, $N$ the number of loops, and $\ell$ the length of the coil. The output capacitance of the GaNFET circuit will generally dominate the capacitive portion of the response, and so the resonator can be treated as a combination of the circuit capacitance of ~1~pF directly to an ideal inductor. 

\begin{equation} \label{eqn:L_ind}
    L_\text{coil} = \frac{{\mu}N^2A}{\ell}.
\end{equation}

For capacitors, we can roughly estimate the capacitance of the structure with a parallel plate approximation using equation \ref{eqn:C_cap}, where $\epsilon$ is the permittivity of the medium, $\ell$ the length of the plates, $w$ the width, and $d$ the separation. If the spacing varies over the  structure, a very rough estimate of the capacitance can be made by using the "average" spacing of the plates over the full structure. The inductance can be over-estimated by forming a loop from the GaNFET to the end of the structure at the average spacing, and using equation \ref{eqn:L_cap} below, where $\mu$ is the permeability of the medium and the other quantities the same as for \ref{eqn:C_cap}.

\begin{equation} \label{eqn:C_cap}
    C_\text{plates} = \frac{{\epsilon}w\ell}{d},
\end{equation}

\begin{equation} \label{eqn:L_cap}
    L_\text{plates} = \frac{{\mu}d\ell}{w}.
\end{equation}

Using these lumped element quantities, we can estimate the "critical" damping ($R_\text{damp}$) needed of any structure using equation \ref{eqn:critDamp}. This takes into account the total capacitance ($C_\text{total}$), inductance ($L_\text{total}$), and resistance ($R_\text{internal})$) of the system. In our circuit, the layout capacitance is approximately 1~pF, the inductance approximately 0.5~nH, and the resistance (from the GaNFET channel and PCB roughness at such high frequencies) approximately \SI{20}{\ohm}, which we add to the quantities calculated for the load above. However, none of these quantities were experimentally verified.  Better estimates of the inductive and capacitive quantities can be found from \cite{Paul2009Inductance:Partial}. Doubling the damping estimated from the rough estimation of these quantities ensured over-damped results.

\begin{equation} \label{eqn:critDamp}
    R_\text{damp} = R_\text{internal} + \sqrt{\frac{4L_\text{total}}{C_\text{total}}}.
\end{equation}

If we would like to go a step further than damping the port voltage and ensure there is no field ringing, we would have to estimate the frequency spectrum of the pulse as well as the locations and magnitudes of the resonances in the load, and then multiply these together in the frequency domain. Calculating the frequency spectrum of the pulse is the most straightforward part of this to estimate, and thus we do it to start. 

We first need to estimate what the port voltage would look like. This can be done in three ways. The first, more accurate approach is to simulate the circuit in LTSpice, using an RLC circuit with the estimated parameters as the load, and inserting the GaNFET of choice. Selecting the simulation time as the repetition rate, interpolating the output to a chosen sample rate, and then taking the Fourier transform of this gives the frequency spectrum. This approach accurately takes into account the GaNFET turn-on time and any coupling from the gate drive circuit. A simpler but less accurate approach for estimating edge shape can be taken by an ideal RLC circuit responding to a step. In this case, it is possible to solve for the response analytically, then take the Fourier transform of this at the resolution and repetition rate that is desired, which we do in the attached code. Third, it should also be possible to analytically solve for the RLC transfer function, then multiply this by an ideal step in the frequency domain to get a fully analytic solution, though we did not do this.


We then must estimate the position of the resonances. We do this only for the capacitive case. We do this by taking the distance from the GaNFET to the point of interest in our system, weighted by the average velocity of propagation in the system. In the simplified case of pure dielectrics ($\mu_r = 1$), we first calculate an effective dielectric constant of the system. This is shown in equation \ref{eqn:effEpsilon}, where $\ell_i$ is the length of each segment, $\ell_\text{tot}$ the total length to the point of interest, and $\epsilon_i$ the permittivity of each segment. Next, we use this effective impedance to derive the position of the first resonance, assuming that this is a quarter-wave resonance (caused by the low impedance at the input port set by the damping resistance and high impedance at the sampling point). This is shown in equation \ref{eqn:effEpsilon}. We estimate the first resonance to be roughly $10^3$ times larger than the DC response, which is consistent with our simulations.

\begin{equation} \label{eqn:effEpsilon}
    \epsilon_\text{eff} \approx \frac{\ell_1}{\ell_\text{tot}}\epsilon_1 + \frac{\ell_2}{\ell_\text{tot}}\epsilon_2 + ...
\end{equation}

\begin{equation} \label{eqn:quarterWave}
    f_\text{res} \approx \frac{4c}{\ell_\text{tot}\sqrt{\epsilon_\text{eff}}}
\end{equation}

We now multiply the frequency response of the driving edge with this resonant point to estimate the total ringing in the field quantities. By tuning the resistance used in our circuit we can filter to whatever ringing accuracy we hope to achieve. If, for example, the first resonance is at 1.4~GHz and we want 1\% ringing, we then would need the amplitude of the driving edge's frequency response at 1.4~GHz to be $10^{-5}$ times that of the DC response. This comes from the need to achieve $10^{-2}$ accuracy in the resonance relative to the ideal step, divided by the enhancement of that frequency component due to the resonance, which we estimate it to be $10^3$. Resonator quality factors vary drastically depending on the metal chosen, surface roughness, dielectric losses, and detailed geometry, and so an improved estimate of this is deeply dependent on the exact structure.

\section{Data Availability}
All data, COMSOL simulations, circuit schematics, and MATLAB scripts for this work are available at

\url{https://github.com/qnngroup/GaN_Pulsers.git}.

\bibliography{references}